\documentclass[final,times,onecolumn]{elsarticle}

\usepackage{pgfplots}
\usepackage{graphicx}
\usepackage{amstext}
\usepackage{amssymb}
\usepackage{amsmath}
\usepackage{textcomp}
\usepackage{color}
\usepackage[T1]{fontenc}
\usepackage{hyperref}
\usepackage{booktabs}
\usepackage{color}
\usepackage{todonotes}
\usepackage{siunitx}

\hypersetup{
    colorlinks=false,
    pdfborder={0 0 0},
}

\biboptions{sort&compress}

\journal{Energy and Buildings}

\begin{document}
\begin{frontmatter}

\title{Performance analysis of heat and energy recovery ventilators using exergy analysis and nonequilibrium thermodynamics}
\author[ntnu1]{Magnus Aa. Gjennestad}
\author[ntnu2]{Eskil Aursand}
\author[ntnu3]{Elisa Magnanelli\corref{cor1}}
\ead{elisa.magnanelli@ntnu.no}
\author[qu4]{Jon Pharoah}

\cortext[cor1]{Corresponding author.}
 \address[ntnu1]{Department of Physics, NTNU - Norwegian University of Science and Technology \\ NO-7491 Trondheim, Norway}
\address[ntnu2]{Department of Energy and Process Engineering,  NTNU - Norwegian University of Science and Technology \\ NO-7491 Trondheim, Norway}
\address[ntnu3]{Department of Chemistry, NTNU - Norwegian University of Science and Technology \\ NO-7491 Trondheim, Norway}
\address[qu4]{Department of Mechanical and Materials Engineering, Queen's University, K7L 3N6 Kingston, Canada}
\begin{abstract}
The increased attention to energy savings has contributed to more
widespread use of energy recovery systems for building ventilation. We
investigate the efficiency of such systems under different outdoor
conditions using exergy analysis and nonequilibrium
thermodynamics. This analysis makes it possible to assess performance
in terms of loss of work potential, to account for the different
quality of energy and to localize and compare the different sources of
loss in the system. It also enables the use of exergy efficiency as a
single performance parameter, in contrast to the several indicators
that are commonly used. These more common indicators are difficult to
compare and relate to each other. Further, since there is no obvious
optimal trade-off between them, it is challenging to combine them and
develop a global performance indicator that allows for a sensible
comparison of different technical solutions and different types of
recovery devices.  We illustrate the concepts by applying the analysis
to a heat recovery ventilator (HRV) and to a structurally similar
membrane energy recovery ventilator (MERV) that can exchange both heat
and moisture. We show how the exergy efficiency can be used to
identify the range of operating conditions for which the recovery
ventilator is not beneficial as the energy cost is greater than the
energy recovery. This is not trivial using traditional performance
parameters, yet it is a natural outcome of exergy analysis. In
addition, we identify the mechanism by which work potential is lost,
which can help the eventual optimization of both the recovery process
and the auxiliary systems present in ventilation systems.

\end{abstract}

\begin{keyword}
 exergy efficiency\sep energy efficiency \sep recovery ventilator\sep ventilation systems \sep nonequilibrium thermodynamics

\vspace{0.2cm} 
\noindent

\textit{doi:} \href{https://doi.org/10.1016/j.enbuild.2018.04.013}{10.1016/j.enbuild.2018.04.013}
 
\vspace{0.2cm}
\noindent

\textcopyright \ 2018. This manuscript version is made available under
the \href{
  http://creativecommons.org/licenses/by-nc-nd/4.0/}{CC-BY-NC-ND 4.0
  license}.

\end{keyword}

\end{frontmatter}

\section{Introduction}
\label{sec:P5_Introduction}
The increased attention to energy savings has led to buildings becoming more heavily insulated~\cite{chwieduk2003towards,sadineni2011passive}. In tightly enveloped buildings, however, air quality becomes an issue as pollutants may rapidly build up to dangerous levels~\cite{walsh1983indoor,jones1999indoor,sakai2004comparison}. In order to ensure good air quality, it has therefore become important to install ventilation systems that exchange indoor air with fresh outdoor  air~\cite{handbook1996hvac,sugarman2005hvac}. 

Temperature and humidity are parameters that play a vital role in the indoor environment and they should be maintained within certain limits to ensure comfort and health~\cite{fanger1970}. This can be a challenge in both cold climates, where outdoor air coming in through the ventilation system can be much colder and drier than is optimal, and in warm climates where the outdoor air may be too moist and warm. Auxiliary heating, air-conditioning, humidification and dehumidification systems are then necessary to compensate for differences between the conditions of the incoming air and the desired indoor conditions.

In recent years, higher demand for indoor comfort, together with an increase in time spent indoors, has caused the energy use of buildings to rise to the same level as the transport and industry sectors~\cite{fehrm2002exhaust,omer2008energy,perez2008review}. This increase has resulted in the installation of heat recovery systems, which are nowadays often required by regulations and codes, in order to reduce building energy consumption~\cite{McCormick2009,standard201362,byggteknisk2016}.

Heat and energy recovery ventilators (HRV/ERV) are devices that transfer heat, or both heat and moisture, from one air stream and deliver it to another~\cite{shurcliff1988air}. Numerous types of recovery ventilators are currently available, having different characteristics and performance~\cite{mardiana2012review}. 

A widespread solution for heat recovery is the flat plate heat exchanger, which is often used due to low capital costs and ease of operation~\cite{mardiana2012review}. In this device, the exhaust and fresh air streams exchange heat across the wall that separates them. In order to obtain large heat exchange surfaces, many parallel channels are usually stacked on top of each other~\cite{kakac2012heat}. A major limitation of this technology is that moisture cannot be transferred between the two air streams. For this reason, Zhang and Jiang~\cite{zhang1999heat} proposed an alternative solution to conventional flat plate heat exchangers that makes use of a water-permeable membrane as  separating wall between the two air streams. In this way, the exhaust and the fresh air can exchange moisture in addition to heat. 

The performance of a HRV/ERV is influenced by several factors. One is the ability to transfer heat from one air stream to the other. As long as the air moisture stays in vapor phase, only sensible heat is exchanged between the streams. However, if the cooled stream reaches the saturation temperature then water condenses~\cite{vasilyev2016modeling} and the latent heat of condensation is released and is available to be transferred to the heated stream.  Another factor is moisture recovery, which can be achieved in membrane-based energy recovery ventilators (MERV). A third factor is the additional fan power required to drive the air flow through the recovery system. While recovering energy from the exhausted air stream, HRVs/ERVs increase the electrical energy required by ventilation fans~\cite{dodoo2011primary} and thus a net energy saving is not guaranteed.

Many works can be found in the literature that assess the performance of recovery ventilators~\cite{zhang1999heat,zhang2000analysis,zhang2002performance,mandegari2009introduction,zhang2010heat, liu2010efficiency,min2010performance,nasif2010membrane,min2011performance,dugaria2015modelling}. Three performance indicators are traditionally considered. The first parameter, $\epsilon_s$, accounts for the recovery of sensible heat and it is often defined as~\cite{liu2017energy}
\begin{equation}
    \label{eq:P5_epsilon_s}
    \epsilon_s=\frac{T^{\text{\emph{ex}}}_{in}-T^{\text{\emph{ex}}}_{out}}{T^{\text{\emph{ex}}}_{in}-T_{0}},
\end{equation}
where $T^{\text{\emph{ex}}}_{in}$ and $T^{\text{\emph{ex}}}_{out}$ are the temperatures of the exhaust air at the inlet and outlet of the recovery ventilator, respectively, and $T_0$ is the temperature of the outdoor ambient. 

If water is condensed on the wall separating the air streams, latent heat is released, and a second performance parameter can be defined as~\cite{min2010performance}
\begin{equation}
    \label{eq:P5_epsilon_h}
    \epsilon_h=\frac{H^{\text{\emph{ex}}}_{in}-H^{\text{\emph{ex}}}_{out}}{H^{\text{\emph{ex}}}_{in}-H_{0}},
\end{equation}
where $H^{\text{\emph{ex}}}_{in}$ and $H^{\text{\emph{ex}}}_{out}$ are enthalpy flows entering and leaving the system with the exhaust air stream, and $H_0$ is the enthalpy flow that enters the ventilator with the fresh air stream.

In systems where water is also exchanged, a third performance indicator is used to describe the recovery of moisture~\cite{liu2017energy}
\begin{equation}
    \label{eq:P5_epsilon_w}
    \epsilon_w=\frac{x_{w,in}^{\text{\emph{ex}}}-x_{w,out}^{\text{\emph{ex}}}}{x_{w,in}^{\text{\emph{ex}}}-x_{w,0}},
\end{equation}
where $x_{w,in}^{\text{\emph{ex}}}$ and $x_{w,out}^{\text{\emph{ex}}}$ are the water mole fractions of the exhaust air at the system inlet and outlet, and $x_{w,0}$ is the water mole fractions in the outdoor air.

While useful by themselves, these indicators are difficult to compare and relate to each other. Since there is no obvious optimal trade-off between them it is challenging to develop a global performance indicator that allows for a sensible comparison of different technical solutions and different types of recovery devices. Moreover, these parameters do not take into consideration the additional need for fan power. Indeed, a fundamental challenge with recovery ventilators is that they involve various energy forms (i.e.\ thermal energy, chemical potential energy, and electric energy), which differ from each other both from a thermodynamic and an economic point of view  (e.g.\ a certain amount of heat at \SI{294}{\kelvin} has less energy quality and it is in principle cheaper to produce than the same amount of electricity). 

In this work, we investigate the efficiency of ventilation recovery systems using exergy analysis and nonequilibrium thermodynamics. This analysis makes it possible to assess the performance of HRV/ERV in terms of loss of work potential (i.e. exergy destruction) or, equivalently, in terms of entropy produced by the process~\cite{kotas2013exergy}. The method allows for the definition of \textit{one single parameter} to quantify the performance of the process, the exergy efficiency~\cite{lior2007energy}, which can account for the quality of different forms of energy. An easier and more meaningful comparison of different technical alternatives is therefore possible. In addition, it is possible to localize and quantitatively compare the different sources of lost work potential in the system, which may aid in design and optimization of the global ventilation system. This approach is in line with a part of the strategy of the American Society of Heating, Refrigerating and Air-Conditioning Engineers (ASHRAE), which has established a technical group named Exergy Analysis for Sustainable Buildings, to promote the use of this concept in the assessment of energy use of buildings~\cite{ashrae}.

In our analysis, we include the losses due to pressure drops in the ventilation recovery system, and thus we account for the additional power consumed by fans. These may contribute significantly to the overall losses~\cite{noh2010effect} but they are often neglected in performance studies and are not reflected in the performance indicators that are used. 

We will consider examples where we calculate and compare the exergy efficiency and the heat and moisture effectiveness of a flat plate HRV and those of a structurally similar MERV. We shall see how ambient conditions influence the efficiency of the two systems in a non-trivial way, that cannot be predicted by the most commonly used heat and moisture effectiveness. We will consider ambient temperatures and relative humidities characteristic of both cold and warm climates.

After presenting the recovery systems in Section~\ref{sec:P5_system}, the thermo-fluid dynamic model developed to describe the recovery ventilators is introduced in  Section~\ref{sec:P5_theory}. In Section~\ref{sec:case}, we present the analyzed cases and  the  most relevant data. The solution procedure and model validation are illustrated in Section~\ref{sec:P5_method}, before results are presented and discussed in Section \ref{sec:P5_results}. Conclusions are summarized in Section~\ref{sec:conclusions}.

\section{System}\label{sec:P5_system}

\begin{figure}[tb]
\centering
\includegraphics[width=0.78\textwidth]{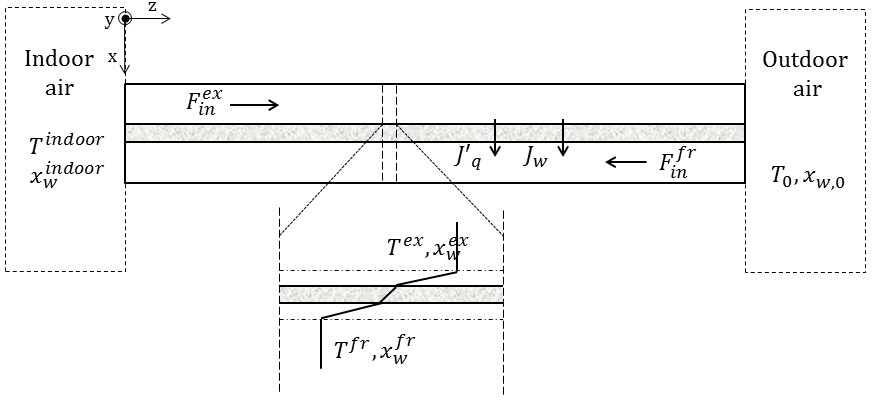}
\caption[width=0.49\textwidth]{Schematic representation of the recovery ventilator. The exhaust air from the indoors flows parallel to the wall that separates the two sides of the system ($z$-direction). The fresh air from the outdoor ambient flows in opposite direction (counter-current configuration). Fluxes of heat, $J'_q$, and water, $J_w$, are  exchanged across the separating wall ($x$-direction).}
\label{fig:P5_scheme}
\end{figure}

Figure~\ref{fig:P5_scheme} shows a schematic representation of a recovery ventilator. The exhaust air from the indoor environment flows from left to right in the positive $z$-direction, while the fresh stream from the outdoor ambient flows parallel to it, but in opposite direction. Since the two streams have different temperatures and composition, driving forces exist for heat and mass transport between them. 

In the thermo-fluid dynamic description of the system, we make the following assumptions.
\begin{itemize}
\item Each air stream can be described as consisting of a bulk region and of two convective boundary layers, one at each side of the walls separating the air streams. In order to enhance heat and mass transfer (and to maintain the structural integrity of the channel), spacers are present in the air channels. These favor turbulence~\cite{woods2013heat}. The bulk of the air streams are therefore considered well mixed in the $x$- and $y$-direction. In the convective boundary layers, however, temperature and concentration gradients are present in the $x$-direction~\cite{cengel2011heat}. 
\item Diffusive fluxes along the $z$-direction are neglected, since they are small in comparison to the advective flow. The diffusive contributions can typically be neglected in systems where the  Peclet number, $\text{Pe}$, is much larger then one~\cite{patankar1980numerical}. In the considered system, we have $\text{Pe}\sim200$. 
\item Humid air is treated as an ideal mixture of ideal gases, where the only components are water and dry air. This assumption is justified by the fact that, in the considered range of temperature and relative humidity, the compressibility factor of humid air at ambient pressure is close to one~\cite{davis1992equation}.
\item The dry air is assumed to be 79\% mole fraction of nitrogen and 21\% oxygen. Other components are present, but they constitute less than 1\% of the gas and are therefore neglected~\cite{brimblecombe1996air}.
\end{itemize}

We consider and compare two structurally similar recovery ventilator solutions. They have the same geometry, but differ in the structural and physical properties of the wall that separates the air streams:
\begin{description}
\item[Heat recovery ventilator (HRV):] The wall separating the air streams is a metal plate. Therefore, only sensible heat can be transferred from one air stream to the other.
\item[Membrane energy recovery ventilator (MERV):] The wall separating the two streams is a water-permeable membrane. Both sensible heat and moisture can be exchanged between the two streams. We assume that water is the only component that can permeate the membrane.
\end{description}
The recovery performances of these systems depends greatly on the amount of exchange surface between the two streams. Therefore, in practical applications, $N$ pairs of exhaust/fresh air channels are stacked to obtain a suitable configuration for the system.

\section{Theoretical formulation}\label{sec:P5_theory}
In Section~\ref{sec:P5_th_balances}, we present the equations used to describe the spatial variation of the system thermodynamic variables. Section \ref{sec:P5_th_transport} deals with heat and mass transport across the system ($x$-direction). The description of transport makes use of the nonequilibrium thermodynamic framework~\cite{Kjelstrup2008}. In Section~\ref{sec:P5_th_efficiencies}, we present the different contributions to the loss of work potential (i.e. exergy destruction), and define the exergy efficiency for the system.

\subsection{Balance equations}\label{sec:P5_th_balances}
The thermodynamic driving forces between the two air streams cause transport of heat and mass from one side to the other. Humid air flows in the ducts, and the water mole fraction varies in space. Therefore, two mass balances are necessary to describe the system, one for the water content of air (subscript $w$) and one for the dry air (subscript $a$). At steady state, the mass balance equations for the exhaust air can be written as:
\begin{eqnarray}
\frac{dF^{\text{\emph{ex}}}_{w}}{dz}&=&-WJ_w\label{eq:P5_dFw}\\
\frac{dF^{\text{\emph{ex}}}_{a}}{dz}&=&-WJ_{a}\quad =\quad 0\label{eq:P5_dFa}
\end {eqnarray}
where $F^{\text{\emph{ex}}}_{w}$ and $F^{\text{\emph{ex}}}_{a}$ are the water and the dry air molar flow rates, $W$ is the width of the duct (in the $y$-direction), and $J_{w}$ and $J_{a}$ are the molar fluxes across the  wall separating the streams. Since the separating wall is impermeable to dry air, $J_a$ equals zero and 
$F^{\text{\emph{ex}}}_{a}$ is constant. The total molar flow rate is given by the sum of the flow rates of the two components, $F^{\text{\emph{ex}}}=F^{\text{\emph{ex}}}_w+F^{\text{\emph{ex}}}_{a}$. 

Since we assume that the diffusive fluxes in the $z$-direction are negligible in comparison to the bulk flow velocity, the component flow rates can be related to the component mole fraction as:
\begin{equation}
F^{\text{\emph{ex}}}_i=x^{\text{\emph{ex}}}_iF^{\text{\emph{ex}}}=x^{\text{\emph{ex}}}_ic^{\text{\emph{ex}}}v^{\text{\emph{ex}}}A^{\text{\emph{ex}}}\label{eq:P5_xi_hot}
\end{equation}
where $x^{\text{\emph{ex}}}_i$ is the mole fraction of the component $i$, $c^{\text{\emph{ex}}}$ is the total molar concentration of gas, $v^{\text{\emph{ex}}}$ is the molar velocity, and $A^{\text{\emph{ex}}}$ is the duct cross sectional area. We use the subscript $i$ to indicate the different components, $i \in \left[w, a\right]$. 

The pressure drop in a duct can be calculated by Darcy--Weisbach equation~\cite{cengel2011heat}:
\begin{equation}
    \frac{dp^{\text{\emph{ex}}}}{dz}=-f^{\text{\emph{ex}}}\frac{\rho^{\text{\emph{ex}}}\left|v^{\text{\emph{ex}}}\right|v^{\text{\emph{ex}}}}{2D^{\text{\emph{ex}}}_h}\label{eq:P5_dp_h}
\end{equation} 
where $f^{\text{\emph{ex}}}$ is the friction factor, $\rho^{\text{\emph{ex}}}$ is the mass density,  $p^{\text{\emph{ex}}}$ is the pressure, and $D^{\text{\emph{ex}}}_h$ is the hydraulic diameter of the duct. 

The energy balance of the exhaust air stream can be written as:
\begin{equation}
\frac{d\left(F^{\text{\emph{ex}}}h^{\text{\emph{ex}}}\right)}{dz}=-WJ_q\label{eq:P5_dFh}
\end{equation}
where $h^{\text{\emph{ex}}}$ is the molar enthalpy of the gas, and $J_q$ is the total heat flux from the exhaust to the fresh air. Applying the chain rule and substituting Eqs.~\ref{eq:P5_dFw} and~\ref{eq:P5_dFa} into Eq.~\ref{eq:P5_dFh}, the energy balance can be rewritten in terms of the molar enthalpy as:
\begin{equation}
F^{\text{\emph{ex}}}\frac{d h^{\text{\emph{ex}}} }{dz}=-W\left(J_q-h^{\text{\emph{ex}}}J\right)\label{eq:P5_Fdh}
\end{equation}
where $J=J_w+J_{a}$ is the total molar flux through the separating wall. Eq.~\ref{eq:P5_Fdh} can be reformulated as an ordinary differential equation (ODE) for the temperature. For an ideal gas, the total enthalpy differential can be written as:
\begin{equation}
dh^{\text{\emph{ex}}}=c_p^{\text{\emph{ex}}}dT^{\text{\emph{ex}}}+\sum_ih_i^{\text{\emph{ex}}}dx^{\text{\emph{ex}}}_i\label{eq:P5_dh}
\end{equation}
where $c_p^{\text{\emph{ex}}}$ is the molar heat capacity, $T^{\text{\emph{ex}}}$ is the temperature, and $h_i^{\text{\emph{ex}}}$ is the partial molar enthalpy. Eq.~\ref{eq:P5_Fdh} can then be rewritten as:
\begin{equation}
c_p^{\text{\emph{ex}}}\frac{dT^{\text{\emph{ex}}}}{dz}=-\frac{W}{F^{\text{\emph{ex}}}}\left(J_q-h^{\text{\emph{ex}}}J\right)-\sum_i h_i^{\text{\emph{ex}}}\frac{dx^{\text{\emph{ex}}}_i}{dz}\label{eq:P5_dT}
\end{equation}
where $M^{\text{\emph{ex}}}$ is the molar mass of the exhaust air.

All the thermodynamic variables introduced above refer to the exhaust air, and they are referred to with the superscript \emph{ex}. We indicate the corresponding variables on the fresh air side with a similar notation and the superscript \emph{fr}. The balance equations for the fresh air are:
\begin{eqnarray}
\hspace{-11mm}\frac{dF^{\text{\emph{fr}}}_{w}}{dz}&=&WJ_w\label{eq:P5_dFw_c}\\
\hspace{-11mm}\frac{dF^{\text{\emph{fr}}}_{a}}{dz}&=&WJ_{a}=0\label{eq:P5_dFa_c}\\
\hspace{-11mm}\frac{dp^{\text{\emph{fr}}}}{dz}&=&-f^{\text{\emph{fr}}}\frac{\rho^{\text{\emph{fr}}}\left|v^{\text{\emph{fr}}}\right|v^{\text{\emph{fr}}}}{2D^{\text{\emph{fr}}}_h}\label{eq:P5_dp_c}\\
\hspace{-11mm}c_p^{\text{\emph{fr}}}\frac{dT^{\text{\emph{fr}}}}{dz}&=&\frac{W}{F^{\text{\emph{fr}}}}\left(J_q-h^{\text{\emph{fr}}}J\right)-\sum_ih_i^{\text{\emph{fr}}}\frac{dx^{\text{\emph{fr}}}_i}{dz}\label{eq:P5_dT_c}
\end {eqnarray}
 
\subsection{Transport between the air streams}\label{sec:P5_th_transport}
Transport between the air streams can be described by use of flux-force relations from nonequilibrium thermodynamics~\cite{Kjelstrup2008}:
\begin{eqnarray}
\Delta\left(\frac{1}{T}\right)&=&r_{q}J_q^{'ex}\label{eq:P5_Jq}\\
-\Delta\left(\frac{\mu_w}{T}\right)+h_w^{\text{\emph{ex}}}\Delta\left(\frac{1}{T}\right)&=&r_{w}J_w\label{eq:P5_J_w}
\end{eqnarray} 
where $\mu_w$ is the chemical potential of water, $r_{q}$ and $r_w$ are the heat and mass transport coefficients, and  $J_q^{'ex}=J_q- J_wh^{\text{\emph{ex}}}_w$ is the measurable heat flux on the exhaust air side. We use the symbol $\Delta$ to indicate the difference in thermodynamic variables between the two sides. The left-hand sides of \mbox{Eqs.~\ref{eq:P5_Jq}-\ref{eq:P5_J_w}} represent the driving forces for heat and mass transport. Coupling between fluxes has been neglected because they are most often small in comparison to the main transport contributions.

Two main phenomena contribute to the transport coefficients. The first contribution is due to the resistance to transport offered by the wall between the two streams. The second contribution is given by the resistance to transport of the convective boundary layers that form in the fluid on each side of the separating wall. The calculation of the transport coefficients is further discussed in \ref{app:P5_A}.

\subsection{Exergy efficiency and loss of useful work}\label{sec:P5_th_efficiencies}
For a recovery ventilator, a task exergy efficiency can be defined as~\cite{kotas2013exergy}:
\begin{equation}
    \label{eq:P5_eta_II}
    \eta=1-\frac{E_{d}}{E_{u}}
\end{equation}
where $E_{u}$ is the exergy of the exhaust air that enters the recovery ventilator (i.e. maximum useful work that can be extracted from the exhaust air stream), and $E_{d}$ is the exergy that is destroyed in the process (i.e. useful work that is lost).     

Two main phenomena contribute to the overall exergy destruction:
\begin{equation}
    \label{eq:P5_W_lost}
    E_{d}=E_{d,irr}+E_{d,amb}
\end{equation}
where $E_{d,irr}$ is the exergy destroyed due to irreversibilities inside the system, and $E_{d,amb}$ is the exergy that is lost by discharging the exhaust air in the outdoor ambient. 

The exergy that is destroyed due to irreversibilities is directly related to the total entropy that is produced during the process, $\Sigma_{irr}$, by the Gouy--Stodola theorem~\cite{gouy1889}:
\begin{equation}
    \label{eq:P5_W_irr}
    E_{d,irr}=T_0\Sigma_{irr}
\end{equation}
where $T_0$ is the temperature of the ambient. The total entropy production of the process is the integral of the local entropy production of a system cross section, $\sigma$, over the system length, $L$~\cite{Kjelstrup2008}:
\begin{equation}
    \label{eq:P5_Sigma}
    \Sigma_{irr}=\int^L_0 \sigma \ dz
\end{equation}
The entropy production can be written as sum of the products of all thermodynamic forces in the system and their conjugated fluxes~\cite{Kjelstrup2008}. Thus, for the present case, the local entropy production is:
\begin{eqnarray}
    \label{eq:P5_sigma}
    \sigma&=&J_q^{'ex}\Delta\left(\frac{1}{T}\right)+J_w\left(-\Delta\left(\frac{\mu_w}{T}\right)+h_w^{\text{\emph{ex}}}\Delta\left(\frac{1}{T}\right)\right)\\
    &&+A^{\text{\emph{ex}}}v^{\text{\emph{ex}}} \left(-\frac{1}{T^{\text{\emph{ex}}}}\frac{dp^{\text{\emph{ex}}}}{dz}\right)+A^{\text{\emph{fr}}}v^{\text{\emph{fr}}}\left(-\frac{1}{T^{\text{\emph{fr}}}}\frac{dp^{\text{\emph{fr}}}}{dz}\right)\nonumber
\end{eqnarray}
where the first term on the right-hand side represents the product between the measurable heat flux and the thermal driving force, and the second is the product between the mass flux and the chemical driving force. The last two terms represent the entropy production due to friction, where the flux is the stream velocity and the viscous flow driving force is $\left(-\frac{1}{T}\frac{dp}{dz}\right)$.

The exergy destroyed due to irreversibilities can also be calculated from the exergy balance on the system, as presented in \ref{app:P5_B_Sigma_irr}. The alternative formulation given by Eq.~\ref{eq:Sigma_irr2} does not require detailed knowledge of the thermodynamic variables across the system, but relies  exclusively on inlet and outlet values of the variables. However, it does not localize where entropy is produced in the system and by what phenomena. This information can be used in designing, improving and optimizing recovery devices. 

The exergy that is lost to the ambient corresponds to the exergy that is destroyed by discharging the exhaust air stream in the ambient (i.e. the exergy of the exhausted air stream when it leaves the recovery ventilator, $E^{\text{\emph{ex}}}_{out}$~\cite{kotas2013exergy}:
\begin{eqnarray}
    \label{eq:P5_W_amb}
    E_{d,amb}&=&E^{\text{\emph{ex}}}_{out}\\
    &=&F^{\text{\emph{ex}}}_{out}\left(h^{\text{\emph{ex}}}_{out}-h^{\text{\emph{ex}}}_{out,0}-T_0\left(s^{\text{\emph{ex}}}_{out}-s^{\text{\emph{ex}}}_{out,0}\right)\right)+F^{\text{\emph{ex}}}_{{out}}RT_0\sum_ix^{\text{\emph{ex}}}_{{i,out}}\ln\left(\frac{x^{\text{\emph{ex}}}_{{i,out}}}{x_{{i,0}}}\right)\nonumber\\
    &=&E_{d,amb,ph}+E_{d,amb,ch}\nonumber
\end{eqnarray}
where $h^{\text{\emph{ex}}}_{out,0}$ and $s^{\text{\emph{ex}}}_{out,0}$ are the enthalpy and the entropy of the gas evaluated at $T_0$ and $p_0$. The subscript $out$  indicates the thermodynamic variable at the outlet of the considered air stream. Here, $E_{d,amb,ph}$ and $E_{d,amb,ch}$ are the physical and the chemical contributions to the exergy lost to ambient. 

The exergy of the exhaust air that enters the recovery ventilator is defined as~\cite{kotas2013exergy}:
\begin{eqnarray}
    \label{eq:P5_W_ideal}
    E_{u}&=&E^{\text{\emph{ex}}}_{out}\\
    &=&F^{\text{\emph{ex}}}_{in}\left(h^{\text{\emph{ex}}}_{in}-h^{\text{\emph{ex}}}_{in,0}-T_0\left(s^{\text{\emph{ex}}}_{in}-s^{\text{\emph{ex}}}_{in,0}\right)\right)+F^{\text{\emph{ex}}}_{in}RT_0\sum_ix^{\text{\emph{ex}}}_{i,in}\ln\left(\frac{x^{\text{\emph{ex}}}_{i,in}}{x_{i,0}}\right)\nonumber
\end{eqnarray}
where $s^{\text{\emph{ex}}}_{in}$ is the molar entropy of the inlet exhaust air,  $h^{\text{\emph{ex}}}_{in,0}$ and $s^{\text{\emph{ex}}}_{in,0}$ are the enthalpy and entropy of the gas evaluated at the temperature and pressure of the ambient (i.e. $T_0$ and $p_0$), and $x_{i,0}$ is the reference composition of the ambient. The subscript $in$ indicates the thermodynamic variable at the inlet of the considered air stream. Similarly to Eq.~\ref{eq:P5_W_amb}, the first term on the right-hand side of Eq.~\ref{eq:P5_W_ideal} represents the physical contribution to the exergy of the exhaust air stream, $E_{u,ph}$, while the second term represents the chemical contribution, $E_{u,ch}$~\cite{kotas2013exergy}. 

By substituting Eq.~\ref{eq:P5_W_irr}, Eq.~\ref{eq:P5_W_amb} and Eq.~\ref{eq:P5_W_ideal} into Eq.~\ref{eq:P5_eta_II}, it is possible to reformulate the expression for $\eta$ as:
\begin{equation}
    \label{eq:P5_eta_ex}
    \eta =  \frac{\left(E^{ex}_{in}-E^{ex}_{out}\right)-E_{d,irr}}{E^{ex}_{in}}
\end{equation}
This reformulation allows us to better understand the meaning of $\eta$. The term at the numerator of Eq.~\ref{eq:P5_eta_ex} represents the net useful work recovered by the recovery ventilator, which is given by the difference between the exergy that is recovered from the exhaust air stream, $\left(E^{ex}_{in}-E^{ex}_{out}\right)$, and the exergy that is dissipated due to irreversibilities, $E_{d,irr}$. 

The denominator in Eq.~\ref{eq:P5_eta_ex} represents the maximum useful work that can be extracted from the exhausted air stream before it is discharged in the ambient. Thus, $\eta$ represents the fraction of the exhaust air exergy that is recovered. In the ideal reversible case where the exergy of the exhaust air is completely recovered with no pressure drop in the system, $\eta$ would be equal to one. The closer  $\eta$ is to one, the larger is the fraction of the exergy available in the exhaust air which is recovered.

Moreover, this parameter allows us to identify those operating conditions for which the recovered exergy, $\left(E^{ex}_{in}-E^{ex}_{out}\right)$, is smaller than the exergy that is destroyed by the ventilator due to irreversibilities, $E_{d,irr}$. For such conditions, $\eta$ is negative. From a thermodynamic point of view, a negative $\eta$ indicates that there exists a better alternative to provide the exergy that is recovered by the recovery ventilator, and it would be better to let the air streams bypass the recovery unit.

\section{Case specification}\label{sec:case}

\subsection{Relevant data}

\begin{table}[tb]
\centering
\caption{Operating conditions and structural and physical parameters assumed in calculations.}
\begin{tabular*}{0.78\textwidth}{@{\extracolsep{\fill} } l c c c }
\toprule
Symbol& Value & Unit & Ref. \\
\hline
& \vspace{-3mm} &  \\
$p_0$ & \SI{e5}{}  &\si{\pascal}&\\
\multicolumn{2}{l}{\textbf{Indoor environment}}&&\\
$T^{indoor}$ & 294 &\si{\kelvin}&\cite{byggteknisk2016}\\
$RH^{indoor}$ &\ 40 &\si{\percent}&\cite{byggteknisk2016}\\
$v_{in}^{\text{\emph{ex}}}$& \SI{0.6}{} &\si{\meter\per\second}&\\
\multicolumn{2}{l}{\textbf{Outdoor ambient}}&&\\
$T_0$ &263-303 &\si{\kelvin}&\\
$RH_0$ &10-90 & \si{\percent}&\\
\multicolumn{2}{l}{\textbf{Geometrical parameters}}&&\\
$L$&0.185&\si{\meter}&\cite{zhang2010heat}\\
$W_{duct}$&0.185&\si{\meter}&\cite{zhang2010heat}\\
$H^{\text{\emph{ex}}}_{duct},H^{\text{\emph{fr}}}_{duct}$&~~\SI{4.0e-3}{}&\si{\meter}&\cite{zhang2010heat}\\
$N$ & 57& -&\cite{zhang2010heat}\\
\hspace{1mm}\textbf{HRV}&&\\
$\delta$&~~\SI{5.0e-4}&\si{\meter}&\\
$\lambda^{wall}$&\SI{2.0e2}{}&\si{\watt\per\meter\per\kelvin}&\\
$P_w$&0& \si{\kilo\mol\per\pascal\per\second\per\meter}&\\
\multicolumn{2}{l}{\textbf{MERV}}&&\\
$\delta$&~~\SI{1.02e-4}{}&\si{\meter}&\cite{zhang2010heat}\\
$\lambda^{wall}$ & \SI{0.13}{} &  \si{\watt\per\meter\per\kelvin}&\cite{zhang2010heat}\\
$P_w$&~~\SI{1.0e-13}{}&\si{\kilo\mol\per\pascal\per\second\per\meter}&\cite{huizing2015selective}\\
\bottomrule
\end{tabular*}
\label{table:P5_data}
\end{table}
The most relevant data used in the calculations are summarized in Table~\ref{table:P5_data}.
We adopt a target indoor relative humidity, $RH^{\text{\emph{ex}}}_{in}$, of \SI{40}{\percent}, and indoor temperature of \SI{294}{\kelvin}, which lie in the range of indoor humidities and temperatures recommended by the Norwegian building code~\cite{byggteknisk2016}. We consider a range of outdoor temperatures and humidity, typical of both cold and warm  climates. 

The geometrical configuration of the two systems is the same, but they differ in structural and physical characteristics of the wall that separates the air streams. The geometrical parameters and the wall properties in Table~\ref{table:P5_data} are used to calculate the transport coefficients in Table~\ref{table:P5_r}. Usually, the air channels contain spacers that improve the structural integrity channels and enhance heat and mass transfer. However, the presence of spacers makes the friction factor and, therefore, the pressure drops larger (Eq.~\ref{eq:P5_dp_h} and Eq.~\ref{eq:P5_dp_c}). We use the relation for friction factor in ducts with spacers determined by the  experimental work in Ref.~\cite{woods2013heat}.

In order to obtain a larger exchange area between the two streams, several air channels are alternately stacked on top of each other to obtain a convenient configuration of the system. Because of symmetry, the complex stacked system can be described by a single pair of exhaust/fresh air channels, where the height of the channel is half of the one of a single channel, $H=H_{duct}/2$, and the width is double that of a single channel multiplied by the number of channel pairs, $W=2W_{duct}N$.

\subsection{Boundary conditions}\label{sec:P5_BC}
In order to solve the set of eight ODEs, we need to impose eight boundary conditions:
\begin{eqnarray}
T^{\text{\emph{ex}}}_{in}&=&T^{indoor}\\
F_{in}^{\text{\emph{ex}}}&=&A^{\text{\emph{ex}}}\frac{v_{in}^{\text{\emph{ex}}}p_0}{RT^{indoor}}\\
F_{w,in}^{\text{\emph{ex}}}&=&F_{in}^{\text{\emph{ex}}}x_{w}^{indoor}\\
p_{out}^{\text{\emph{ex}}}&=&p_0\\
T^{\text{\emph{fr}}}_{in}&=&T_0\\
F_{a,in}^{\text{\emph{fr}}}&=&F_{a,in}^{\text{\emph{ex}}}\\
F_{w,in}^{\text{\emph{fr}}}&=&F_{a,in}^{\text{\emph{fr}}}\frac{x_{w,0}}{1-x_{w,0}}\\
p_{out}^{\text{\emph{fr}}}&=&p_0
\end{eqnarray}
The boundary condition values are calculated from the data reported in Table~\ref{table:P5_data}. In addition, we consider that the dry air stream extracted from the indoor environment needs to be the same as the dry air stream supplied from the outdoor ambient.

\section{Solution procedure}\label{sec:P5_method}

With the given assumptions, the system is properly described by a set of eight ordinary differential  equations (Eqs.~\ref{eq:P5_dFw}-\ref{eq:P5_dFa}, Eq.~\ref{eq:P5_dT}, Eq.~\ref{eq:P5_dp_h}, and Eqs.~\ref{eq:P5_dFw_c}-\ref{eq:P5_dp_c}). Since the boundary conditions are specified at both ends of the system~(Section~\ref{sec:P5_BC}), the problem is a two-point boundary value problem. The problem is solved by use of the MATLAB \texttt{bvp4c}-solver, which is a collocation solver that implements the three-stage Lobatto IIIa formula and bases the mesh selection and error control on the residual of the solution~\cite{shampine2000solving}. 

\section{Model validation}
\subsection{Consistency check}
In order to check the consistency of the thermodynamic framework presented in Section~\ref{sec:P5_theory}, the exergy destruction due to irreversibility has been calculated for all investigated cases according to both Eq.~\ref{eq:P5_W_irr} and Eq.~\ref{eq:Sigma_irr2}. This allows us to check that neither the first law or second law is violated anywhere inside the system. The relative error between the results given by the
two equations are order of the numerical accuracy of the calculations (10$^{-5}$).

\subsection{Comparison with empirical correlations}
Heat and moisture exchange effectiveness, $\epsilon_{s}$ and $\epsilon_{w}$, are two of the parameters that are most commonly used  in the literature to characterize recovery performance. Most published research uses heat and moisture effectiveness as a criterion to verify the accuracy of simulations as well as of experiments~\cite{liu2017energy}. Widely used empirical correlations exist to calculate the heat and moisture effectiveness based on the design parameters of the HRV/MERV (i.e. the number of transfer unit for  transport of heat and moisture, $NTU_s$ and $NTU_w$)~\cite{kays1984compact,zhang2002effectiveness}. In order to verify the accuracy of the model, the computed heat and moisture effectiveness were compared to those predicted by the $\epsilon$-$NTU$ relations that presented in References~\cite{kays1984compact} and~\cite{zhang2002effectiveness}. The results agree within \SI{1.6}{\percent} (with the exception of a narrow range of outdoor conditions where $\epsilon_s$ diverges), indicating that the model gives a good description of a recovery ventilator.

\subsection{Sensitivity analysis}
A sensitivity analysis has been carried out on four parameters, to show how the exergy efficiency varies as a function of the selected parameters. The result of the sensitivity analysis (reported in \ref{app:P5_C_Sensitivity}) shows that uncertainty in the input parameters does not significantly affect the results and it does not reveal any unexpected relationships between inputs and outputs.

\section{Results}\label{sec:P5_results}

\subsection{Transport between streams}

\begin{table}[tb]
\centering
\caption{Calculated transport coefficients for the HRV and for the MERV, at $z=L/2$. Calculations are carried out at $T_0= \SI{263}{\kelvin}$ and $RH_0=\SI{60}{\percent}$.}
\begin{tabular*}{0.78\textwidth}{@{\extracolsep{\fill} } l c c c}
\toprule
 & HRV &MERV & Unit\\
\hline
& \vspace{-3mm} &  \\
$r_q^{h,conv}$ & \SI{1.4e-7}{} & \SI{1.4e-7}{} & \si{\meter\squared\second\per\joule\per\kelvin}\\
$r_q^{wall}$ & ~\SI{2,7e-11}{} & \SI{8.8e-9}{} & \si{\meter\squared\second\per\joule\per\kelvin}\\
$r_q^{c,conv}$ & \SI{1.5e-7}{} & \SI{1.5e-7}{} & \si{\meter\squared\second\per\joule\per\kelvin}\\
$r_w^{h,conv}$ & -& \SI{2.6e6}{}~~ & \si{\meter\squared\second\joule\per\kelvin\per\kilo\mol\squared}\\
$r_w^{wall}$  & - & \SI{4.5e9}{}~~ & \si{\meter\squared\second\joule\per\kelvin\per\kilo\mol\squared}\\
$r_w^{c,conv}$ &- & \SI{2.6e6}{}~~ & \si{\meter\squared\second\joule\per\kelvin\per\kilo\mol\squared}\\
\bottomrule
\end{tabular*}
\label{table:P5_r}
\end{table}

Two main phenomena contribute to the  overall heat and mass transport coefficients. One contribution is due to the heat transport resistance of the wall material, while a second contribution depends on transport across the convective  boundary layers adjacent to the separating  wall. Table~\ref{table:P5_r} presents the different contributions to the total heat and mass transport coefficients in the HRV and MERV as calculated according to the equations presented in \ref{app:P5_A}. The convective layers are responsible for most of the heat transport resistance. In the HRV, the convective resistances to heat transport are  four orders larger than that due to conduction across the separating wall. This justifies the common practice of neglecting the wall conductive resistance ~\cite{liu2017energy} for this case. Membrane materials for MERV typically have a lower thermal conductivity than the metallic materials in HRV (three orders of magnitude lower, in the present case). The  resistance to conductive heat transport is small in comparison to the convective resistance even in the case of the MERV, contributing only \SI{3.5}{\percent} to the total resistance. 

In contrast to the case of heat transport, the main resistance to mass transport is caused by diffusion through the membrane material. In the present case, the convective resistance is only \SI{0.1}{\percent} of the total resistance to mass transport. Convective transport may however  play a more important role under different operating conditions as other studies have reported convective contributions to water mass transport as high as \SI{30}{\percent} for different flow regimes~\cite{min2011performance}.

\begin{figure}[tb]
\centering
\includegraphics[width=0.78\textwidth]{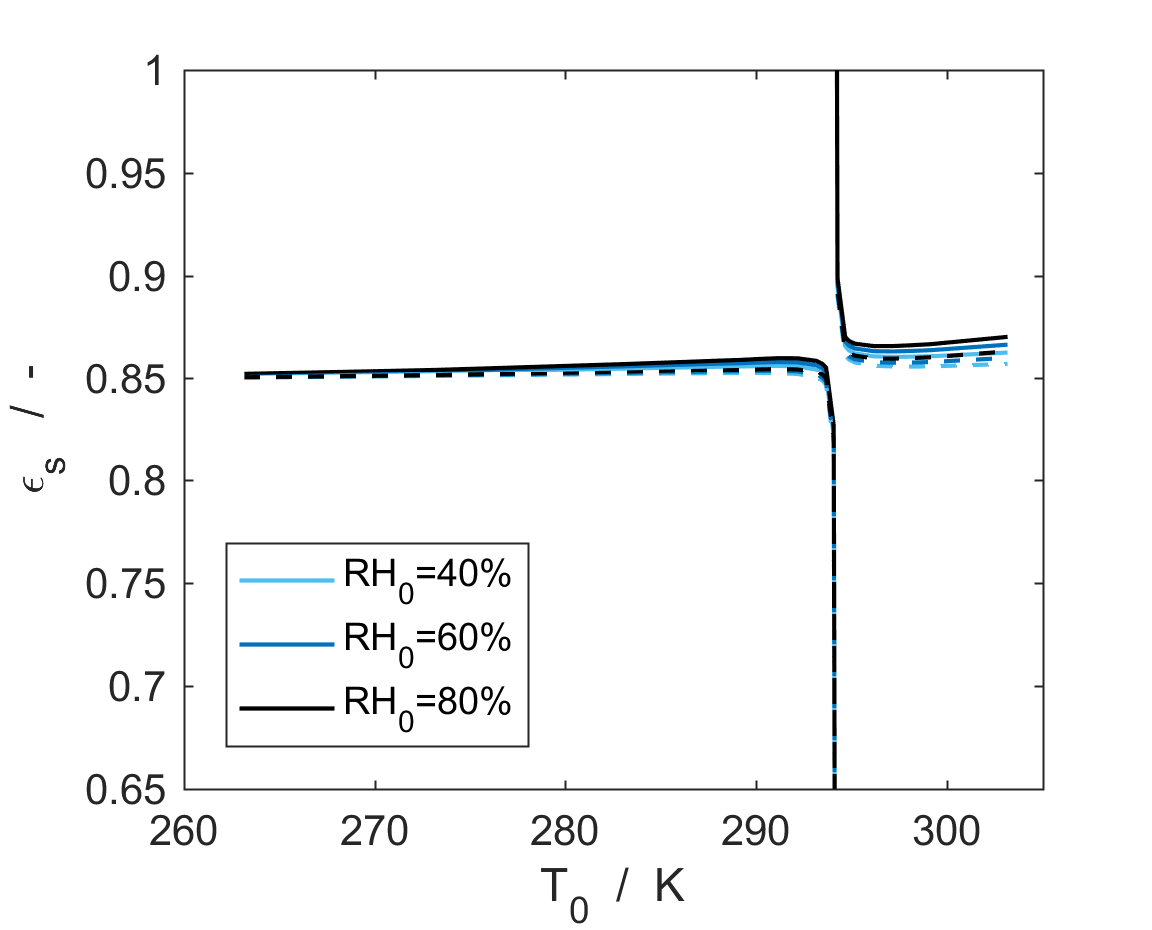}
\caption[width=0.48\textwidth]{Sensible heat exchange effectiveness, $\epsilon_{s}$, for the HRV (solid lines)  and for the MERV (dashed lines), for $RH_0=\SI{40}{\percent}$ (light blue lines), $RH_0=\SI{60}{\percent}$ (blue lines), and $RH_0=\SI{80}{\percent}$ (black lines). At $T_0=T^{indoor}$, $\epsilon_{s}$ is not defined.}
\label{fig:P5_epsilon_s}
\end{figure}

\begin{figure}[tb]
\centering
\includegraphics[width=0.78\textwidth]{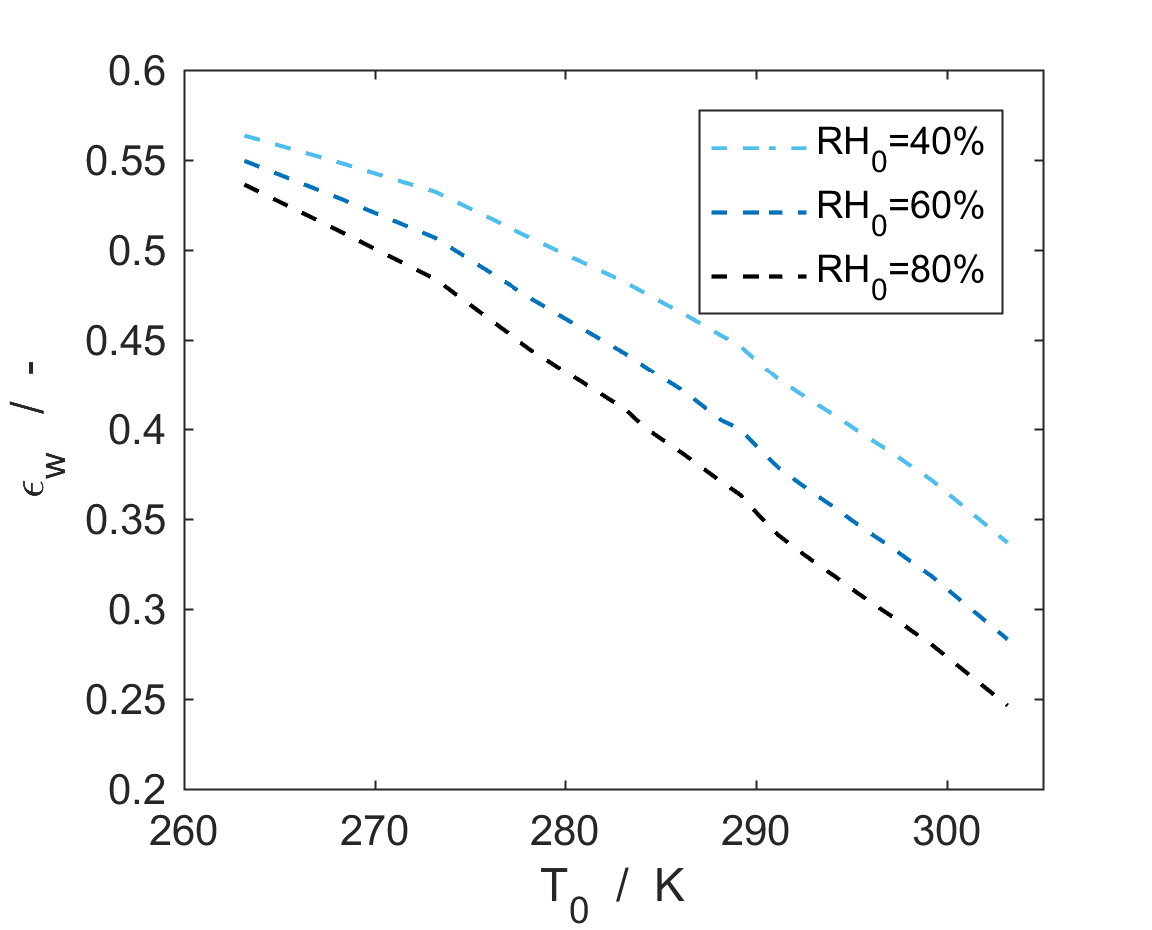}
\caption[width=0.48\textwidth]{Moisture exchange effectiveness, $\epsilon_{w}$ for the MERV (dashed line) as functions of the outdoor temperature, for $RH_0=\SI{40}{\percent}$ (light blue line), $RH_0=\SI{60}{\percent}$ (blue line), and $RH_0=\SI{80}{\percent}$ (black line). Since HRV does not exchange any water, $\epsilon_{w}$ is everywhere zero for the HRV. At $x_{w,0}=x_w^{indoor}$, $\epsilon_{w}$ is not defined.}
\label{fig:P5_epsilon_w}
\end{figure}
\begin{figure}[tb]
\centering
\includegraphics[width=0.78\textwidth]{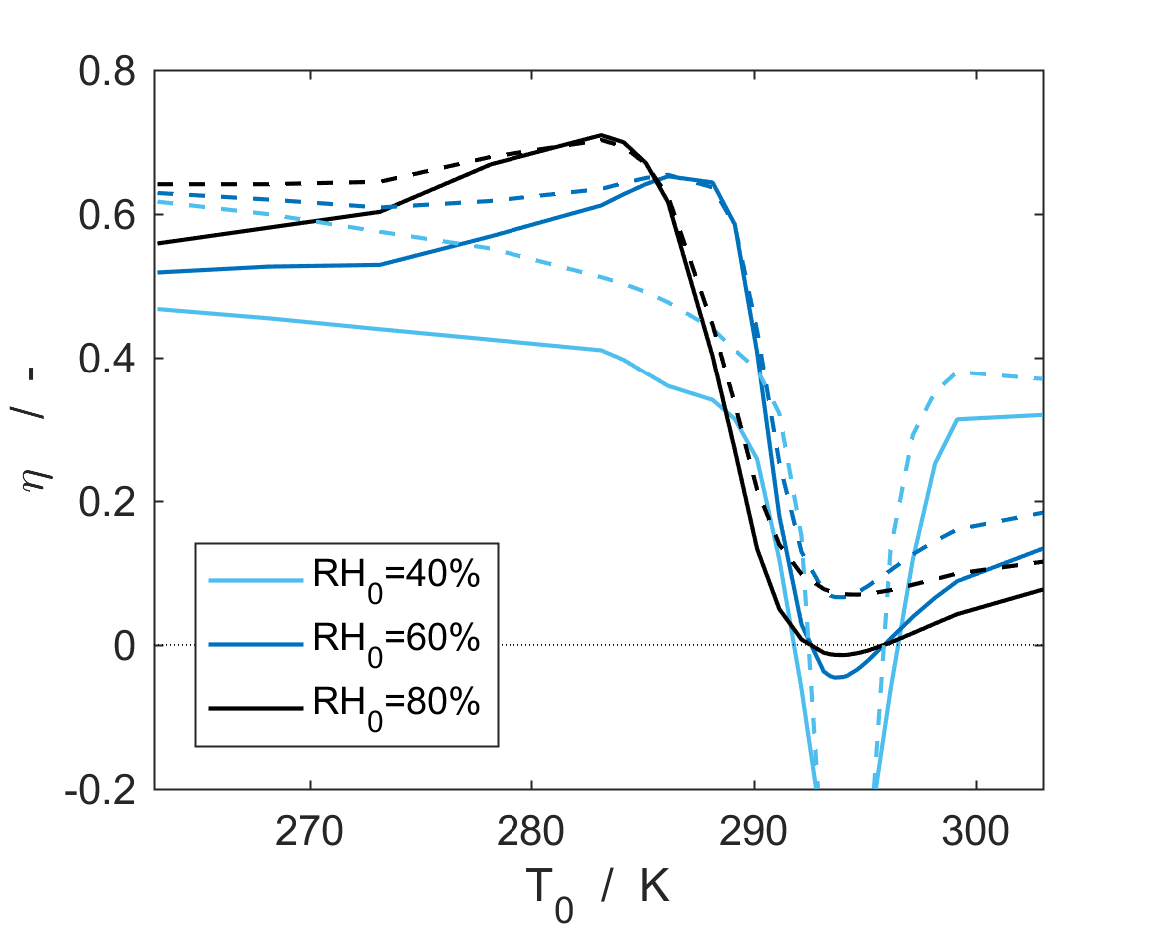}
\caption[width=0.48\textwidth]{Exergy efficiency, $\epsilon_{\text{ex}}$, as defined by Eq.~\ref{eq:P5_eta_II} for the HRV (solid lines) and for the MERV (light blue lines) as functions of the outdoor temperature, for $RH_0=\SI{40}{\percent}$ (red lines), $RH_0=\SI{60}{\percent}$ (blue lines), and $RH_0=\SI{80}{\percent}$ (black lines).}
\label{fig:P5_eta}
\end{figure}

\begin{figure}[tb]
\centering
\includegraphics[width=0.78\textwidth]{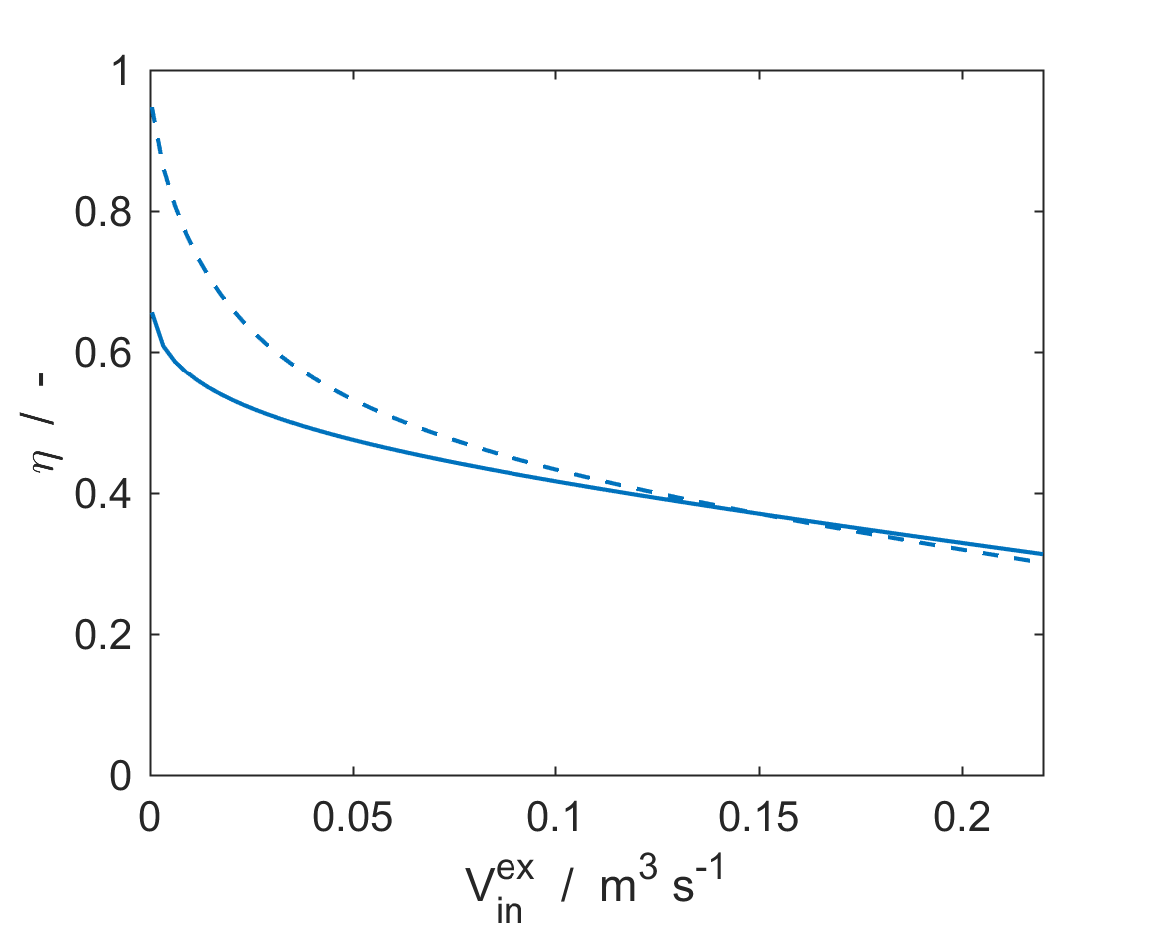}
\caption[width=0.48\textwidth]{Exergy efficiency, $\eta$, for the HRV (solid line) and for the MERV (dashed line) as functions of the volume flow rate. Calculations are carried out at $T_0=\SI{263}{\kelvin}$ and $RH_0=\SI{60}{\percent}$.}
\label{fig:P5_eta_V}
\end{figure}

\begin{figure}[tbp]
\centering
\includegraphics[width=0.78\textwidth]{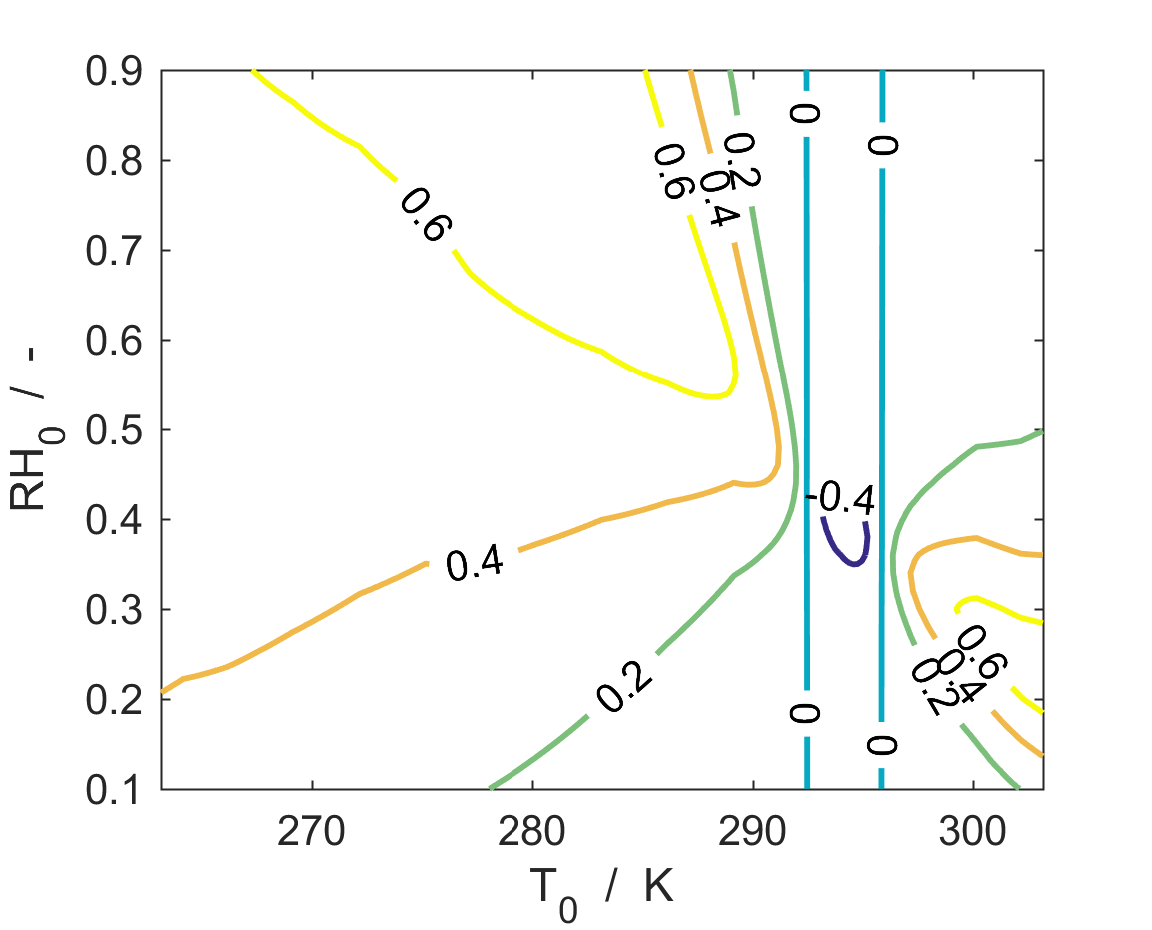}
\caption[width=0.48\textwidth]{Exergy efficiency, $\eta$, for the HRV as function of the outdoor ambient temperature and of the outdoor relative humidity.}
\label{fig:eta_iso_HRV}
\end{figure}
\begin{figure}[tbp]
\centering
\includegraphics[width=0.78\textwidth]{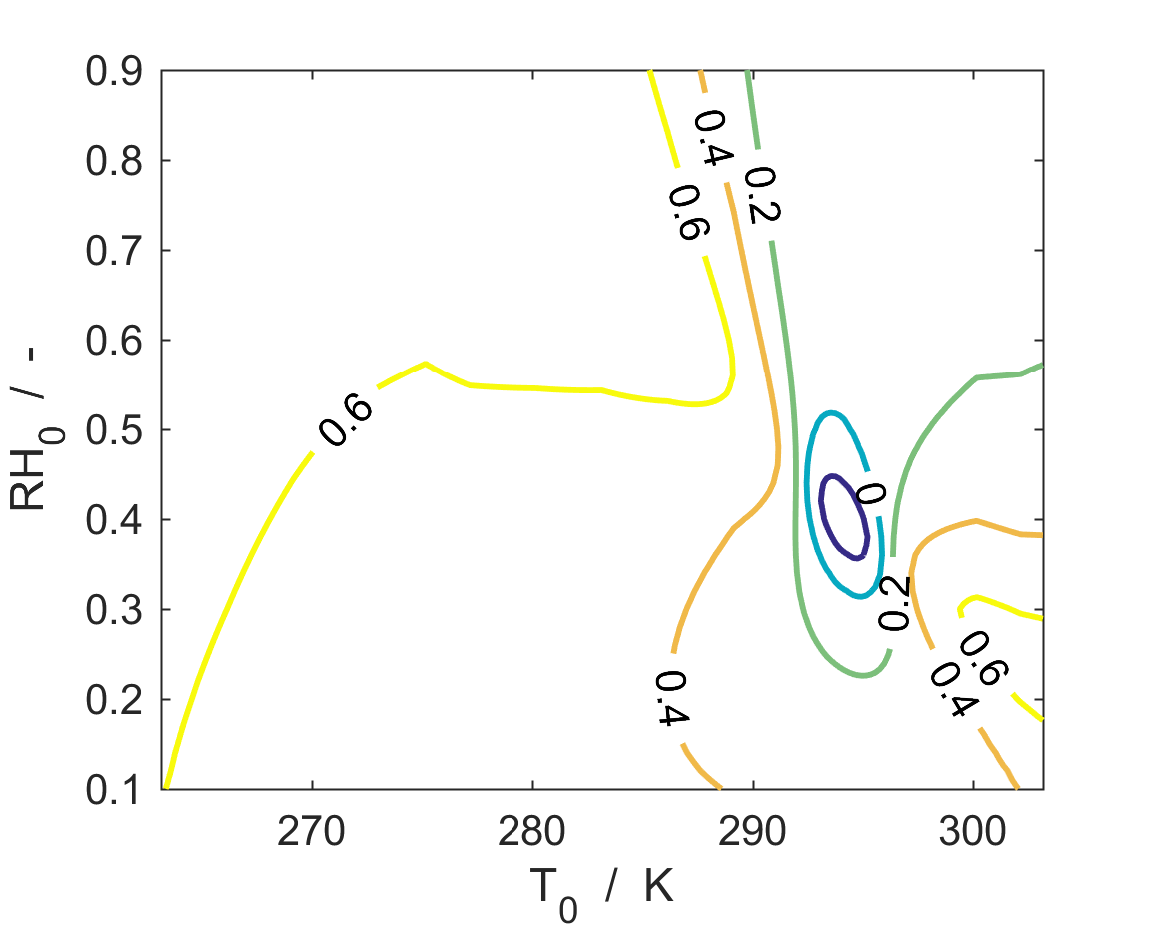}
\caption[width=0.48\textwidth]{Exergy efficiency, $\eta$, for the MERV as function of the outdoor ambient temperature and of the outdoor relative humidity.}
\label{fig:eta_iso_MERV}
\end{figure}

\begin{figure}[tb]
\centering
\includegraphics[width=0.78\textwidth]{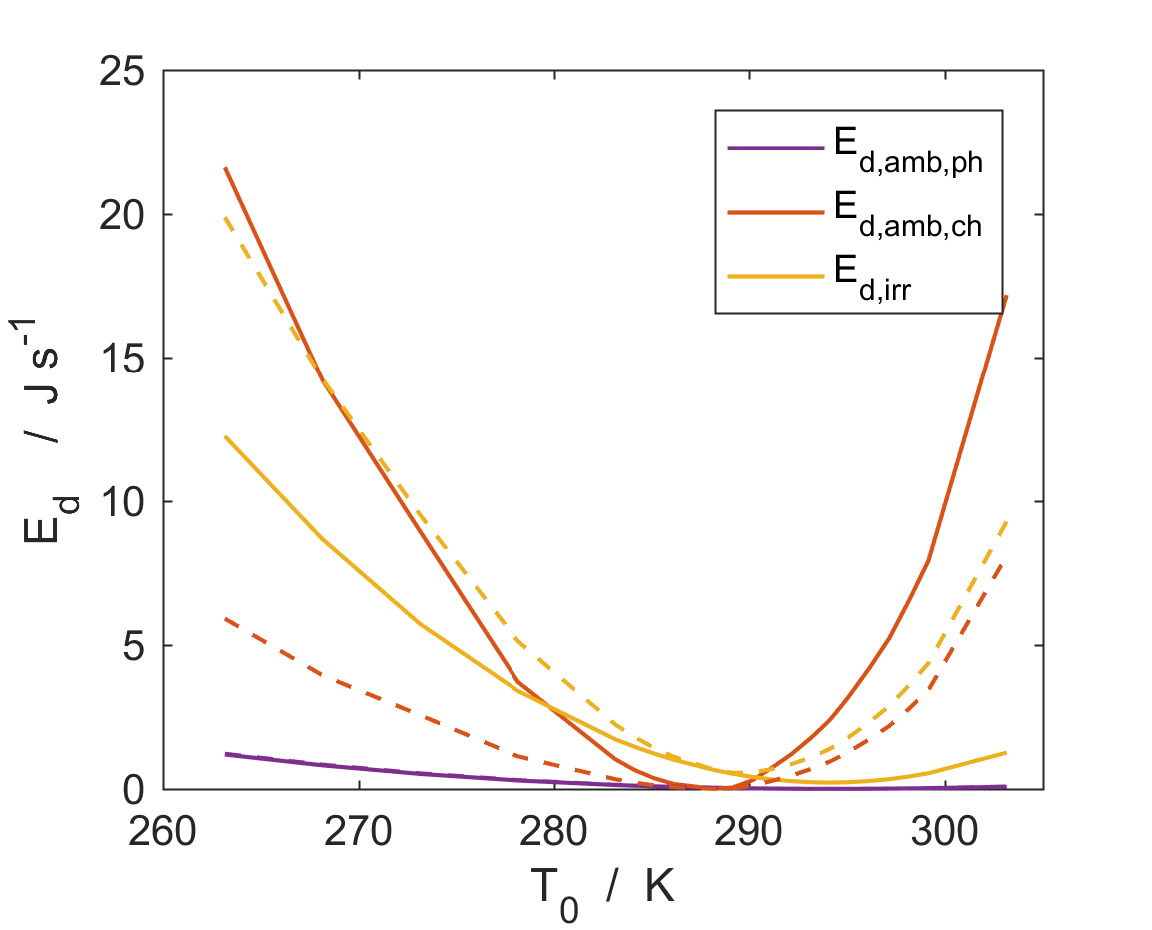}
\caption[width=0.48\textwidth]{Physical useful work lost to the ambient, $W_{amb,ph}$ (purple lines), chemical useful work lost to the ambient, $W_{amb,ch}$ (orange lines), and useful work lost to irreversibilities, $W_{irr}$ (yellow lines), for the HRV (solid lines) and for the MERV (dashed lines) as functions of the outdoor temperature. The solid and dashed purple lines ($W_{amb,ph}$) coincide at the scale of the representation. Calculations are carried out at $RH_0=\SI{60}{\percent}$.}
\label{fig:P5_W_lost}
\end{figure}

\subsection{Heat and moisture effectiveness and exergy efficiency}\label{sec:eta}

Figure~\ref{fig:P5_epsilon_s} shows the sensible heat exchange effectiveness of the HRV (solid lines) and of the MERV (dashed lines) calculated according to Eq.~\ref{eq:P5_epsilon_s}. The sensible effectiveness is very similar for the HRV and the MERV for all considered outdoor temperatures and relative humidities. The slightly higher $\epsilon_{s}$ for the HRV is due to the lower resistance to heat transport offered by the metal wall that separates the streams. The sensible effectiveness does not have a strong dependence on the outdoor conditions, and increases only slightly for higher outdoor temperatures. Since the denominator in Eq.~\ref{eq:P5_epsilon_s} is zero when the outdoor and indoor temperatures are equal (at $T_0=T^{indoor}$), $\epsilon_s$ cannot be defined for this case. Because of friction, a small increase of the exhausted stream temperature occurs even when indoor and outdoor temperatures are equal. This phenomenon makes $\epsilon_{s}$ diverge for outdoor temperatures close to the indoor temperature. 
 
Figure~\ref{fig:P5_epsilon_w} represents the moisture transfer effectiveness, $\epsilon_w$, for the MERV (dashed lines) as a function of the outdoor temperature, for $RH_0=\SI{40}{\percent}$ (light blue line), $RH_0=\SI{60}{\percent}$ (blue line), and $RH_0=\SI{80}{\percent}$ (black line). In the HRV, the air streams do not exchange any moisture, and this parameter is always zero. For the MERV, $\epsilon_w$  decreases with decreasing  outdoor temperature. Similar behaviour has  been observed by others (e.g. the experimental work in Reference~\cite{min2011performance}). When the water mole fractions of the indoor and outdoor air are equal $\epsilon_w$ is not defined.

Figure~\ref{fig:P5_eta} illustrates the exergy efficiency, as defined by Eq.~\ref{eq:P5_eta_II}, of the HRV (solid lines) and the MERV (dashed lines). The behavior of $\eta$ as a function of the ambient temperature is very complex, however it is similar for different outdoor relative humidities. In cold climates, for $RH_0=\SI{60}{\percent}$ (blue line), and $RH_0=\SI{80}{\percent}$ (black line), $\eta$ of the HRV increases as the outdoor temperature increases, until it reaches the value for which indoor and outdoor water mole fraction are the same,  $x^{indoor}_w=x_{w,0}$. For a further increase of $T_0$, $\eta$ drops. When $T^{indoor}=T_0$, the exergy efficiency of the HRV is  negative, since useful work is lost, due to pressure drops, without generating any useful output. For $T_0$ higher  than the indoor temperature, $\eta$ increases with increasing ambient temperature. When $x^{indoor}_w=x_{w,0}$, the air streams cannot exchange moisture in any  of the configurations, and the efficiencies of the HRV and MERV are approximately the same. For  $RH_0=\SI{40}{\percent}$ (light blue lines), $x^{indoor}_w$ and $x_{w,0}$ are the same when $T^{indoor}=T_0$. Under such conditions, no useful work can be recovered from the exhausted stream, and the exergy efficiency is not defined. 

For the MERV, $\eta$ is roughly constant both in cold and warm climates, and it drops from one value to the other in the temperature interval between the conditions for which $x^{indoor}_w=x_{w,0}$ and $T^{indoor}=T_0$. The comparison between the two systems shows that the exergy efficiency of the MERV is never smaller than that of the HRV. In cold climates, the difference between the two efficiencies increases with decreasing temperature. For $RH_0=\SI{60}{\percent}$ (blue dashed line) and $RH_0=\SI{80}{\percent}$ (black dashed line), the $\epsilon_{\text{ex}}$ of the MERV is never negative. However, for temperatures and water fraction of the outside air equal to those of the indoor space, the MERV exergy efficiency is negative (light blue dashed line), indicating that no net benefit is obtained from the recovery ventilator. Instead, it is detrimental to the performance of the ventilation system as electric energy is expended to compensate for pressure drop. Under such conditions it would be better to let the air streams bypass the recovery unit.

Figure~\ref{fig:P5_eta_V} shows how the exergy efficiency of the HRV (solid line) and of the MERV (dashed line) varies as a function of the exhaust air volume flow rate. The air flow rate is varied within a range that allows for the air velocity to stay  under \SI{5}{\meter\per\second}, in order to avoid unnecessary noise generation. The exergy efficiency is largest when the flow rate approaches zero, and it decreases for increasing flow rates. There is thus a trade-off between the amount of air that can be treated by a ventilator of a given size and the resulting exergy efficiency. A larger HRV/MERV will retain a high efficiency for larger flow rates, but would require larger investment costs. In practice, the air volume flow rate is determined by the size of the building and by the rate at which the indoor air needs to be replaced. The latter is often dictated by local regulations, e.g.~\cite{byggteknisk2016}. With the flow rate given, one may choose an acceptable $\epsilon_{\text{ex}}$ and increase the size of HRV/MERV, or change other design parameters, until the desired efficiency is achieved. 

This is one example of a way in which the exergy efficiency may guide the design and sizing of such devices. Nowadays, graphs picturing heat exchange effectiveness as a function of volume flow rate are used to dimension recovery ventilators. By switching to the use of a graphs like the one in Fig.~\ref{fig:P5_eta_V}, it is possible to dimension the recovery ventilators taking into account not only the effect of heat recovery, but also the effect of moisture recovery and pressure drops. 

Figure~\ref{fig:P5_eta_V} shows that for high flow rates the exergy efficiency of the HRV is higher than that of the MERV. This is due to the fact that the mass transport resistance is large in comparison to that of heat transport. Therefore, the transport of heat is more facile than the mass transport. Accordingly, for high air velocities, the water recovery is very low, and heat recovery represents the dominant contribution to the exergy efficiency. The resistance to convective heat transport decreases with increasing flow rate and, hence the wall heat transport resistance becomes relatively more important. Since the wall thermal resistance of the MERV is two orders of magnitude larger than that of the HRV, the MERV heat recovery and exergy efficiency are smaller than those of the HRV.

Figures~\ref{fig:eta_iso_HRV} and~\ref{fig:eta_iso_MERV} show the exergy efficiency as function of both outdoor temperature and relative humidity. These figures are useful e.g.~to identify the range of outdoor conditions for which the recovery ventilator is responsible for increasing the energy need of the building (area enclosed by the light blue lines where $\eta$ is zero). Under such conditions, it is more beneficial to bypass the HRV/MERV. The comparison between the two pictures shows that the range of unfavorable outdoor conditions is much smaller for the MERV than for the HRV. Moreover, the MERV retains a exergy efficiency larger than \SI{60}{\percent} in a larger portion of the outdoor condition space.

In the present case, pressure losses are low, and thus the operational range for which $\eta$ is negative is small. However, under different operating conditions and configurations (e.g.\ higher air velocities or larger friction coefficients inside the ducts), such losses could have larger influence on the overall performance, and enlarge the range of operating conditions for which the use of a HRV/MERV is not beneficial. 

If the indoor ambient conditions are to be maintained constant, the exergy of the building needs to stay constant over time. This means that the exergy which is not recovered by the recovery ventilator, needs to be supplied to the indoor ambient by other auxiliary systems (i.e. heaters/air coolers, humidifiers/dehumidifiers, and fans), whose energy consumption one needs to pay for. In an ideal case, $\eta$ would be equal to one and none of the auxiliary systems would be necessary. For a fixed set of operating conditions, the higher $\eta$ is, the lower is the exergy load that the auxiliary systems need to supply for. When $\eta$ is negative, the exergy load that the auxiliary systems need to supply is larger than that they would need to provide if the recovery ventilator was not used. 

The comparison between $\epsilon_{s}$, $\epsilon_{w}$ and the exergy efficiency $\eta$ shows that these three parameters have very different behaviors, and that the combination of $\epsilon_{s}$, $\epsilon_{w}$ into a single performance indicator is not straightforward. Performance analysis based on $\epsilon_{s}$ and $\epsilon_{w}$ can therefore create difficulties in the comparison of different solutions, especially when alternatives where only heat is recovered need to be compared with solutions where both heat and moisture are recovered.

Figure~\ref{fig:P5_W_lost} shows for $RH_0=\SI{60}{\percent}$ the contributions to exergy destruction due to physical (purple lines) and chemical (orange lines) work potential that is lost to the ambient, and due to the work potential lost due  to irreversibilities in the system (yellow lines), for the HRV (solid lines) and for the MERV (dashed lines). Similar results are obtained for the other considered outdoor relative humidities (not reported here).  

The physical lost work (or physical exergy destruction, $E_{d,amb,ph}$) is approximately the same for the two systems (solid and dashed purple lines coincides at the plot accuracy), since the outlet temperature of the exhaust air is about the same in the HRV and MERV. Moreover, $E_{d,amb,ph}$ is quite small in comparison to the other work losses, especially at very low and at very high outdoor temperatures.

On the other hand, $E_{d,amb,ch}$ is much larger in the HRV than in the MERV, since none of the moisture in the exhaust air is recovered in the HRV and all the  chemical potential work of the inlet exhaust air is lost.

The loss of work potential due to irreversibilities ($E_{d,irr}$) in the MERV is always larger than in the case of the HRV. Indeed, $E_{d,irr}$ strongly depends on the irreversible heat and mass transfer and on the pressure drops in the system (see Eq.~\ref{eq:P5_sigma}). Since the contributions due to heat transfer and pressure drops are approximately the same in the two systems, the difference between $E_{d,irr}$ is mainly due to the additional contribution of moisture transport in the MERV. Figure~\ref{fig:P5_W_lost} shows that $E_{d,irr}$ goes through a minimum. For the HRV, minimum $E_{d,irr}$ is obtained for $T^{indoor}=T_0$. Under these conditions, no heat is exchanged between the air streams, and the  only contribution to $E_{d,irr}$ is due to the pressure drops.  In the MERV, the minimum in $E_{d,irr}$ is located between the outdoor conditions for which the sensible heat transfer is zero ($T^{indoor}=T_0$), and the condition for which the moisture transfer is zero ($x^{indoor}_w=x_{w,0}$).  

The overall exergy destruction in the system gives an indication of the energy that one needs to supply to the indoor ambient to maintain temperature and humidity levels that are compatible with comfortable indoor conditions. In order to maintain comfort, the ventilation system is normally coupled with auxiliary systems that compensate for the work lost during ventilation. In particular, auxiliary systems add or remove sensible heat from the indoor air (e.g. radiators or air conditioners), others control the moisture level (e.g. dehumidifiers or humidifiers), while fans compensate for the pressure drops in the ventilation system. The knowledge on the different contributions to the lost work allows us to locate and compare the ideal amount of energy needed by the different auxiliary systems to maintain the desired temperature and humidity levels. Since different auxiliary systems might be supplied by different energy sources, this approach makes it easier to include considerations on the primary energy use of buildings.

Figure~\ref{fig:P5_W_lost} shows that, for the HRV, most of the energy that needs to be supplied to the system is used for moisture control, both in cold and warm climates. With the use of the MERV in cold climates, the energy used for such purpose is reduced to a level similar to that needed for temperature control. This is not the case for warm climates. For ambient temperatures close to that of the indoor temperature, the use of the HRV is not beneficial, since it increases the fan power, without bringing any recovery of heat or moisture. On the other hand, due to moisture exchange, the use of the MERV remains beneficial for ambient temperatures close to the indoor temperature, when the indoor and outdoor water content is different.

\section{Conclusions}\label{sec:conclusions}

We have used exergy analysis and nonequilibrium thermodynamics to characterize the performance of heat and energy recovery ventilators. We considered and compared the performances of a heat recovery ventilator (HRV) and those of a structurally similar membrane energy recovery ventilator (MERV). This kind of analysis made it possible to quantify, assess and compare the performances of HRV/MERV in terms of a single parameter, the exergy efficiency. This parameter describes the loss of work potential and may account for all different sources of work loss and for differences in energy quality. 

Traditional effectiveness parameters defined by Eqs.~\ref{eq:P5_epsilon_s}-\ref{eq:P5_epsilon_w} are most commonly used in the literature to characterize recovery ventilators performance.  While useful by themselves, these effectiveness parameters are difficult to compare and relate to each other. Since there is no obvious optimal trade-off between them, it is challenging to combine them in a way that allows for a sensible comparison of different technical solutions such as the HRV and MERV.

We showed that the exergy efficiency can be used to identify the range of operating conditions for which the recovery ventilator is responsible for increasing the energy need of the building, and for which it is more beneficial to bypass the HRV/MERV. When the exergy efficiency is negative, the recovered heat and moisture are not enough to compensate for the fan power necessary to drive the air flow in the recovery unit. Such phenomenon cannot trivially be predicted by traditional performance parameters. 

Moreover, by using graphs that represent exergy efficiency as a function of air volume flow rate, it is possible to dimension the recovery ventilators taking into account not only the effect of heat recovery, but also the effect of moisture recovery and  pressure drops.

The exergy analysis additionally enabled  the localization and comparison of the various components of loss of useful work in the systems. This means that the analysis can be used not only to assess the efficiency of the process, but also to draw considerations on  the design of the auxiliary devices used in the process.

\section*{Acknowledgments}
The Norwegian University for Science and Technology, the Research Council of Norway (project number 262644) and the VISTA program, a basic research program in collaboration with The Norwegian Academy of Science and Letters and Statoil, are acknowledged for financial support.

\appendix
\renewcommand{\thefigure}{A.\arabic{figure}}
\setcounter{figure}{0}
\section{Transport coefficients}\label{app:P5_A}
The  transport coefficients have two main contributions. A first contribution is due to the  resistance to transport offered by the wall separating the two streams, $r_i^{wall}$. A second contribution is  given by the resistance offered by the convective boundary layers that are present at the two sides of the separating wall, $r_i^{h,conv}$ and $r_i^{c,conv}$. Similarly to resistances in series, the overall transport coefficients can be written as:
\begin{equation}
r_i=r_i^{h,conv}+r_i^{wall}+r_i^{c,conv}\label{eq:r_i}
\end{equation}
\subsection{Heat transport coefficients}
The wall heat transport coefficient can be derived from experimental data on the thermal conductivity of the wall material. Thermal conductivity experiments are typically carried out at zero mass flux. According to Fourier's law, the measurable heat flux is:
\begin{equation}
J'_q=-\lambda^{wall} \frac{\Delta T}{\delta}\label{eq:Fourier}
\end{equation}
where $\lambda^{wall}$ is the thermal conductivity of the wall material, and  $\delta$ is the wall thickness. By  comparing Eq.~\ref{eq:Fourier} and Eq.~\ref{eq:P5_Jq}, we can relate the wall heat transport coefficient to the experimental thermal conductivity of the wall material:
\begin{equation}
r_q^{wall}=\frac{\delta}{\lambda^{wall} T^2}
\end{equation}

The heat transport resistance due to the convective boundary layer can be calculated from the convective heat transfer coefficient, $h^{conv}$, which is defined as:
\begin{equation}
    J'_q=-h^{conv} \Delta T
\end{equation}
where $\Delta T$ is the difference in temperature between the membrane surface and the gas bulk.
The coefficient depends on the characteristics of the flow, and it is related to the Nusselt number by the following relation:
\begin{equation}
h^{conv}=\text{Nu}\frac{\lambda}{D_h}\label{eq:Jq_conv}
\end{equation}
where $\lambda$ is the thermal conductivity of the fluid.
Empiric relations exist that relate the Nusselt number to the characteristics of the flow. In order to maintain the channel geometry and to enhance heat and mass transfer, spacers are usually present in the air channels. In this work, we use the correlations for Nusselt number that have been established for flow in ducts with  spacers~\cite{woods2013heat}:
\begin{equation}
    \label{eq:Nu}
    \text{Nu}=j\text{Re}\text{Pr}^{1/3}
\end{equation}
where $j$ is the Colburn heat transfer factor, $\text{Re}$ is the Reynolds number, and $\text{Pr}$ is the Prandtl number.
The Colburn heat transport factor is determined experimentally as a function of the Reynolds number~\cite{woods2013heat}:
\begin{equation}
    \label{eq:j}
    j=\frac{C_0}{{\text{Re}}^m}
\end{equation}
where $C_0$ and $m$ are two fitted parameters.

By comparing Eq.~\ref{eq:Jq_conv} with Eq.~\ref{eq:P5_Jq}, we can relate the convective heat transport resistance to the convective heat transport coefficient:
\begin{equation}
r_q^{conv}=\frac{1}{h^{conv}T^2}
\end{equation}

\subsection{Mass transport coefficients}
Mass transport through semi-permeable membranes is usually characterized by permeability parameters. The permeability of a material to a component is determined experimentally at isothermal conditions (i.e. $\Delta T$=0). In the present case, the membrane permeability to water, $P_w$, is defined as:
\begin{equation}
J_w=-P_w\frac{\Delta p_w}{\delta}\label{eq:Ji_wall}
\end{equation}
where  $\Delta p_w$ is the water partial pressure difference between the two sides of the membrane. 
By comparing Eq.~\ref{eq:Ji_wall} with Eq.~\ref{eq:P5_J_w}, we can relate the membrane mass transport resistance to the permeability:
\begin{equation}
r_w^{wall}=\frac{\delta}{TP_w}\frac{\Delta \mu_w}{\Delta p_w}
\end{equation}

The convective resistance to water transport can be calculated from the convective mass transport coefficient,  $k^{conv}$, which is defined as~\cite{cengel2011heat}:
\begin{equation}
J_w=-k^{conv}\Delta c_w\label{eq:Ji_conv}
\end{equation}
where $c_w$ is the water molar concentration.
For air-water vapor mixtures, the convective mass transport coefficient can be expressed with good accuracy by the Lewis relation~\cite{cengel2011heat}:
\begin{equation}
k^{conv}=\frac{h^{conv}}{c_pc}
\end{equation}
where $c$ and $c_p$ are the molar concentration and the molar heat capacity of humid air. 
The comparison of Eq.~\ref{eq:Ji_conv} with Eq.~\ref{eq:P5_J_w} gives:
\begin{equation}
r_w^{conv}=\frac{1}{Tk^{conv}}\frac{\Delta \mu_w}{\Delta c_w}
\end{equation}

\renewcommand{\thefigure}{B.\arabic{figure}}
\setcounter{figure}{0}
\section{Exergy destruction calculations from the entropy balance on the system}\label{app:P5_B_Sigma_irr}
In a steady state, the exergy destruction  of a process due to internal irreversibilities  can also be calculated from the exergy balance on the system~\cite{kotas2013exergy}:
\begin{eqnarray}
    \label{eq:Sigma_irr2a}
    E_{d,irr}&=& \left(E_{in}^{\text{\emph{ex}}}+E_{in}^{\text{\emph{fr}}}\right)-\left(E_{out}^{\text{\emph{ex}}}+E_{out}^{\text{\emph{fr}}}\right)\\
    &=&\left[\left(H_{in}^{\text{\emph{ex}}}+H_{in}^{\text{\emph{fr}}}\right)-\left(H_{out}^{\text{\emph{ex}}}+H_{out}^{\text{\emph{fr}}}\right)\right]- T_0\left[\left(S_{in}^{\text{\emph{ex}}}+S_{in}^{\text{\emph{fr}}}\right)-\left(S_{out}^{\text{\emph{ex}}}+S_{out}^{\text{\emph{fr}}}\right)\right]\nonumber
\end{eqnarray}
where $E_{in}^{\text{\emph{ex}}}$ and $E_{in}^{\text{\emph{fr}}}$ are the exergy streams entering the recovery ventilator with the exhaust air stream and the fresh air stream respectively, and  $E_{out}^{\text{\emph{ex}}}$ and $E_{out}^{\text{\emph{fr}}}$ are the exergy streams leaving the system with the exhaust air stream and the fresh air stream respectively. Here, $H$ and $S$ are enthalpy and  entropy flow respectively.

By considering that, due to energy conservation, the sum of the enthalpy streams entering the system needs to be equal to the sum of those leaving the system, then Eq. \ref{eq:Sigma_irr2a} reduces to:
\begin{eqnarray}
    \label{eq:Sigma_irr2}
    E_{d,irr}&=& T_0\left[\left(S_{out}^{\text{\emph{ex}}}+S_{out}^{\text{\emph{fr}}}\right)-\left(S_{in}^{\text{\emph{ex}}}+S_{in}^{\text{\emph{fr}}}\right)\right]\\
    &=&T_0\left[\left(F_{out}^{\text{\emph{ex}}}s_{out}^{\text{\emph{ex}}}+F_{out}^{\text{\emph{fr}}}s_{out}^{\text{\emph{fr}}}\right)-\left(F_{in}^{\text{\emph{ex}}}s_{in}^{\text{\emph{ex}}}+F_{in}^{\text{\emph{fr}}}s_{in}^{\text{\emph{fr}}}\right)\right]\nonumber
\end{eqnarray}
With this formulation, $E_{d,irr}$ can be calculated from the knowledge of flow rates, temperature, composition, and pressure at the inlet and outlet of the system only. Therefore, we do not need any additional information than the one needed to calculate the effectiveness parameters according to Eqs.~\ref{eq:P5_epsilon_s}-\ref{eq:P5_epsilon_w}. 

\renewcommand{\thefigure}{C.\arabic{figure}}
\setcounter{figure}{0}

\section{Sensitivity analysis}\label{app:P5_C_Sensitivity}

\begin{figure}[tb]
\centering
\includegraphics[width=0.78\textwidth]{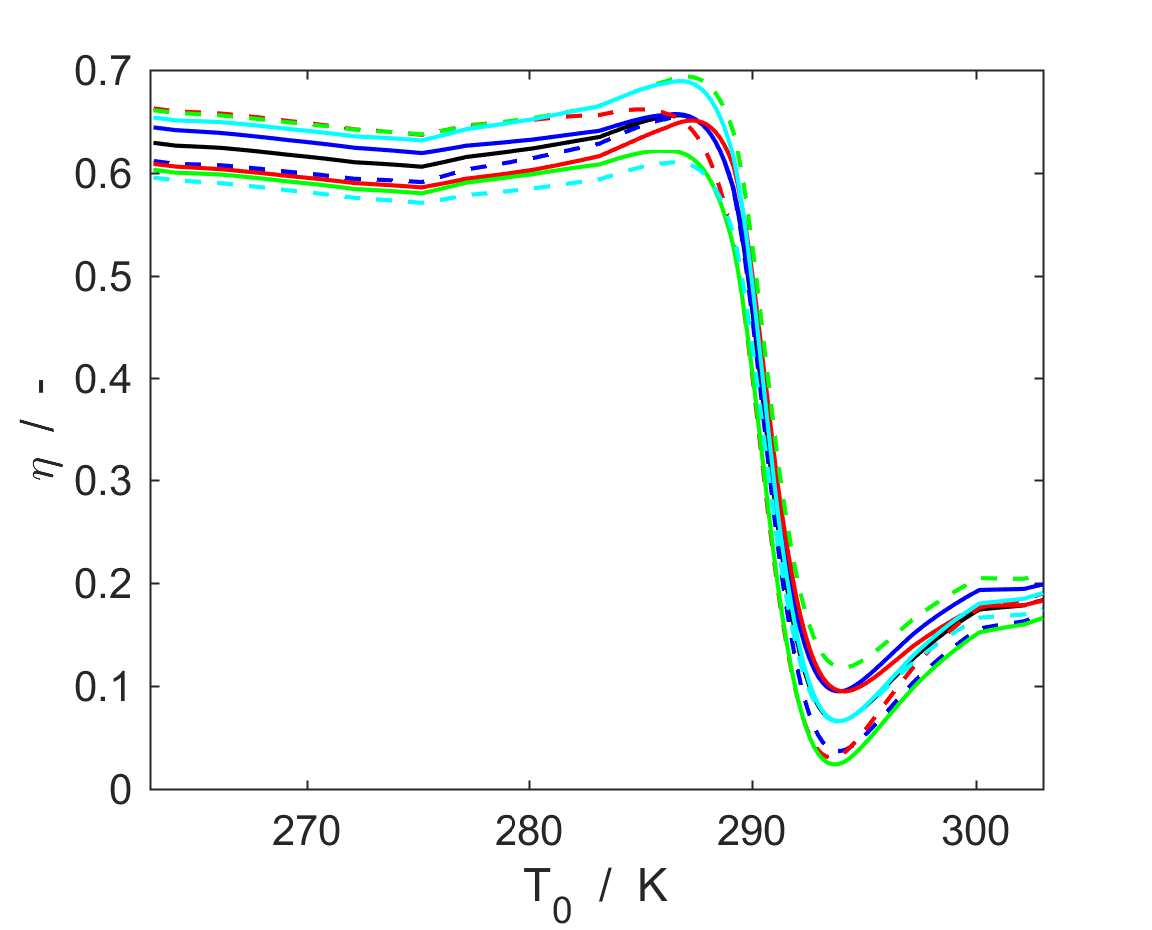}
\caption[width=0.48\textwidth]{Exergy efficiency, $\eta$, of the MERV as a function of the outdoor temperature, for variation of the considered input parameter of \SI{+20}{\percent} (solid lines) and \SI{-20}{\percent} (dashed lines). The considered input parameters are duct height (red lines), membrane permeability (blue lines), inlet air velocity (green lines), and convective heat transport coefficient (light blue lines). The black line represent the results for unvaried input parameters. Calculations are carried out at $T_0=\SI{263}{\kelvin}$ and $RH_0=\SI{60}{\percent}$. }
\label{fig:eta_sens}
\end{figure}
The sensitivity analysis is carried out with a dual purpose. First, the analysis allows us to test the sensitivity of the results to uncertainties in the input parameters. Second, it enables us to obtain a deeper understanding of the relationships between input parameters and results and to possibly reveal any errors in the model, if unexpected behaviors should appear. The input parameters considered in the analysis are duct height, $H_{duct}^{ex}=H_{duct}^{ex}$, membrane permeability, $P_w$, and inlet velocity of the exhausted air, $v_{in}^{ex}$, and convective heat transport coefficient, $h_{conv}$, which are varied from \SI{-20}{\percent} to \SI{+20}{\percent} of their original value reported in Table~\ref{table:P5_data} or calculated according to Eq.~\ref{eq:Jq_conv}.

Figure~\ref{fig:eta_sens} reports the results of the sensitivity analysis for the exergy efficiency. The variation of the input parameters has only a modest impact on $\eta$. The maximum variation of $\eta$ at 263~K is \SI{-5.2}{\percent} (obtained for \SI{-20}{\percent} variation of $h_{conv}$), while it is \SI{-9.3}{\percent} at \SI{303}{\kelvin} (obtained for \SI{+20}{\percent} variation of the inlet air velocity).


\end{document}